\documentclass[aps,prx,twocolumn,10pt,amsfonts,amssymb,showpacs,superscriptaddress,floatfix,longbibliography]{revtex4-2}

\usepackage[utf8]{inputenc}
\usepackage{graphicx}
\usepackage{dcolumn}
\usepackage{bm}
\usepackage{subfigure}
\usepackage{multirow}
\usepackage{amsmath}
\usepackage{color}
\usepackage{hyperref}
\usepackage{wasysym}
\usepackage{mathrsfs}
\usepackage[english]{babel}

\usepackage{times}

\begin{document}

\setlength{\baselineskip}{0.4cm}\addtolength{\topmargin}{1.5cm}

\title{OFC-like Behavior in Experimental Granular Piles}

 \author{K. Duplat}
 \affiliation{Universit\'e Lyon 1, CNRS, Institut Lumi\`ere Mati\`ere, UMR5306, F-69622 Villeurbanne, France.}

 \author{A. Douin}
 \affiliation{Universit\'e Lyon 1, CNRS, Institut Lumi\`ere Mati\`ere, UMR5306, F-69622 Villeurbanne, France.}

 \author{E. Saurety}
 \affiliation{Universit\'e Lyon 1, CNRS, Institut Lumi\`ere Mati\`ere, UMR5306, F-69622 Villeurbanne, France.}

 \author{G. Simon}
 \affiliation{Universit\'e Lyon 1, CNRS, Institut Lumi\`ere Mati\`ere, UMR5306, F-69622 Villeurbanne, France.}

 \author{E. Altshuler}
 \affiliation{Henri Poincar\'e" Group of Complex Systems, Physics Faculty, University of Havana, 10400 Havana, Cuba.}
 
 \author{K. J. M{\aa}l{\o}y}
 \affiliation{Department of Physics, University of Oslo, P.O.B. 1048, Blindern N-0316, Oslo, Norway.}

 \author{O. Cochet-Escartin}
 \affiliation{Universit\'e Lyon 1, CNRS, Institut Lumi\`ere Mati\`ere, UMR5306, F-69622 Villeurbanne, France.}

\author{O. Ramos}
\email{osvanny.ramos@univ-lyon1.fr}
 \affiliation{Universit\'e Lyon 1, CNRS, Institut Lumi\`ere Mati\`ere, UMR5306, F-69622 Villeurbanne, France.}

\date{\today}

\begin{abstract}

Scale-invariant avalanche dynamics are commonly associated with criticality and robust, universal size exponents. Dissipation is generally expected to drive the system away from the critical point, progressively suppressing large events. The Olami--Feder--Christensen (OFC) model challenges this picture: in its non-conservative regime, the avalanche-size exponent is non-universal and can exceed the mean-field value $\tau=3/2$, while system-spanning events persist even at large dissipation. Here, we show that these apparently anomalous features are also observed experimentally in a two-dimensional granular system displaying scale-invariant avalanche dynamics. By increasing interparticle friction, and therefore dissipation, the avalanche-size exponent increases from $\tau=1.58$ to $\tau=1.83$, while the upper cutoff remains proportional to the system size. We further identify similarities between experiment and the OFC model in their memory effects, local dynamics, and the emergence of better-than-random predictability of large events. The latter indicates that the system does not remain permanently critical, but instead evolves through configurations with different propensities to generate extreme events. Our results suggest that OFC-like dynamics are not merely an anomalous feature of a particular model, but may provide a relevant framework for understanding scale-invariant dynamics with $\tau>3/2$ in real dissipative systems.

\end{abstract}

\maketitle

\section{Introduction}\label{intro}

From the Ising model \cite{Budrikis2024} and the Game of Life \cite{gardner1970life} to the sandpile model at the origin of Self-Organized Criticality (SOC) \cite{BTW1987}, cellular automata have long served as simple yet powerful models in statistical physics. Their appeal lies in the fact that complex collective behavior can emerge from minimal microscopic rules, making them useful proxies for studying phase transitions and non-equilibrium phenomena.

Within the framework of SOC, a central objective is to explain the emergence of scale-free avalanche dynamics. In such systems, activity occurs in bursts whose sizes follow a power-law distribution of the form $P(s)\sim s^{-\tau}$, indicating the absence of a characteristic scale. This type of behavior has been reported in a wide range of slowly driven systems, including granular avalanches in piles \cite{Frette1996, Held1990, Rosendahl1993, Altshuler2001, Ramos2009}, and faults \cite{Lherminier2019, Houdoux2021, Daniels2008, Zadeh2019a}, subcritical fracture \cite{Maloy2006, Xu2019, Baro2013, Stojanova2014, Bares2018}, seismology \cite{Gutenberg1956, Bak1989, Main1996, Kawamura2012}, superconducting vortices \cite{Altshuler2004}, and even neuroscience \cite{Beggs2003} and financial markets \cite{Bouchaud2024}. 

Among these phenomena, earthquakes represent one of the most familiar and extensively studied examples of scale-invariant avalanche dynamics. The empirical Gutenberg-Richter law, which describes the power-law distribution of earthquake magnitudes, has therefore become a central reference point for interpreting scale-free behavior in driven dissipative systems \cite{Duplat2025}. In this context, the Olami-Feder-Christensen (OFC) earthquake model \cite{OFC1992} was introduced as a minimal cellular automaton designed to capture essential ingredients of fault dynamics while remaining analytically and computationally tractable. Since its introduction, the OFC model has been widely studied as a paradigm for earthquake-like avalanche dynamics within the broader framework of scale-free dynamics \cite{Hergarten2002, Helmstetter2004, Peixoto_Prado2006, Ramos2006, Ramos2010, DEARCANGELIS2016}. However, its relevance extends beyond seismology, as it also provides a general framework for exploring avalanche statistics in dissipative  systems \cite{Duplat2026}.

 One of the most distinctive properties of the OFC model is the nonuniversal character of its avalanche size exponent. In contrast to many critical models \cite{Alstrom1988, BTW1987, Zapperi_1995, Chessa1999, Dhar_1999, Fisher19998, LeDoussal2009}, where exponents take robust universal values, although limited to $\tau \leq 3/2$,  in the OFC model $\tau$ depends on the level of dissipation in the dynamics. This sensitivity to dissipation makes the OFC model a particularly appealing candidate for describing a broader class of scale-free dynamics, especially in regimes where $\tau > 3/2$, as observed in many real dissipative systems, including earthquakes \cite{Navas2019, Duplat2025} and granular faults \cite{Lherminier2019, Houdoux2021}.

Whether this distinctive dependence of the exponent on dissipation is merely a peculiarity of the model, with little relevance for real systems, or instead reflects a generic property of dissipative avalanche dynamics remains an open question. In this work, we address this issue by studying a classical granular pile \cite{Ramos2009}. We show that its avalanche statistics follow the same trend as in the OFC model when the level of dissipation is varied.

\section{Experimental setup}\label{Exp}

\begin{figure}[t!]
\includegraphics[ width=3.3in]{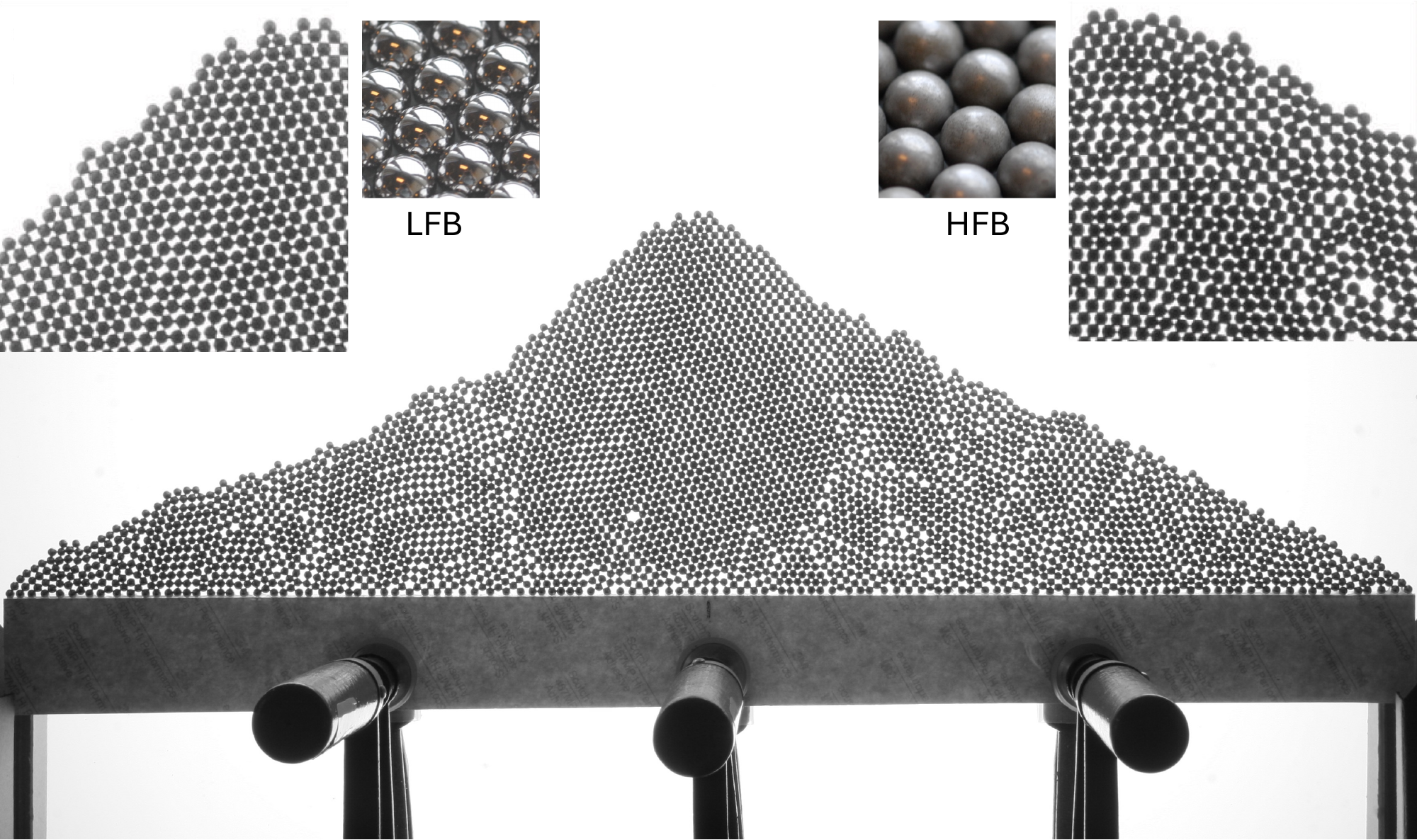}
\caption{Image of a typical pile with low friction beads (LBF). Notice the polycrystalline structure of the system. Left: Details of the apex of the pile and zoom revealing the polished nature of the LFB. Right: Details of the apex of a typical pile obtained with high friction beads (HFB) and zoom on the grains. A more disordered structure appears.}
\label{fig_setup} 
\end{figure}

The experimental setup (Fig.~\ref{fig_setup}) is the same used in \cite{Ramos2009}. The base consists of a 60-cm-long row of $4 \pm 0.005$ mm steel spheres separated by random spacings of 0, 1, 2, or 3 mm \cite{Altshuler2001}. The beads are glued onto an acrylic surface and confined between two parallel vertical glass plates separated by 4.5 mm, forming a quasi-two-dimensional geometry. Identical steel beads are released one by one from a height of 28 cm above the center of the base, leading to the formation of a granular pile. The boundaries of the base are open, allowing beads to leave the pile.

After each bead is added, the system is allowed to relax for a few seconds to ensure that all motion within the pile has ceased. The pile is then imaged using a Canon D20 digital camera with a spatial resolution of 21 pixels per bead diameter, after which the next bead is released. Each experiment contains more than 55,000 bead additions and lasts more than 310 hours. To avoid transient effects during the pile growth, the first 5,000 events are not considered for statistical analysis.

Image processing is used to determine the positions of the centers of all beads. The avalanche size $s$ is defined as the number of beads that move between two consecutive bead additions. A bead is considered to have moved if, in the subsequent image, no bead center is found within a distance smaller than or equal to one seventh of the bead diameter from its previous position. This definition may fail in the rare case where a bead is replaced exactly by another one at the same position after an avalanche. However, visual inspection shows that such events are extremely uncommon, since avalanches typically modify the local packing disorder, making exact replacements statistically negligible.

Two main experiments under different dissipation values were performed: Experiment 1 was conducted using low-friction beads (LFB, Fig.~\ref{fig_setup}) with a Coulomb friction coefficient $f=0.1$. This corresponds to the original dataset reported in \cite{Ramos2009}, which we reanalyze here. The beads were subsequently treated with a diluted {\it aqua regia} solution (1/4 $NO_{3}$ + 3/4 $HCl$), increasing the friction coefficient to $f=0.4$. To investigate the effect of increased dissipation, Experiment 2 was performed under the same conditions as Experiment 1 but using these high-friction beads (HFB, Fig.~\ref{fig_setup}).

\section{Results}\label{Res1}

\begin{figure}[b!]
\includegraphics[ width=3.3in]{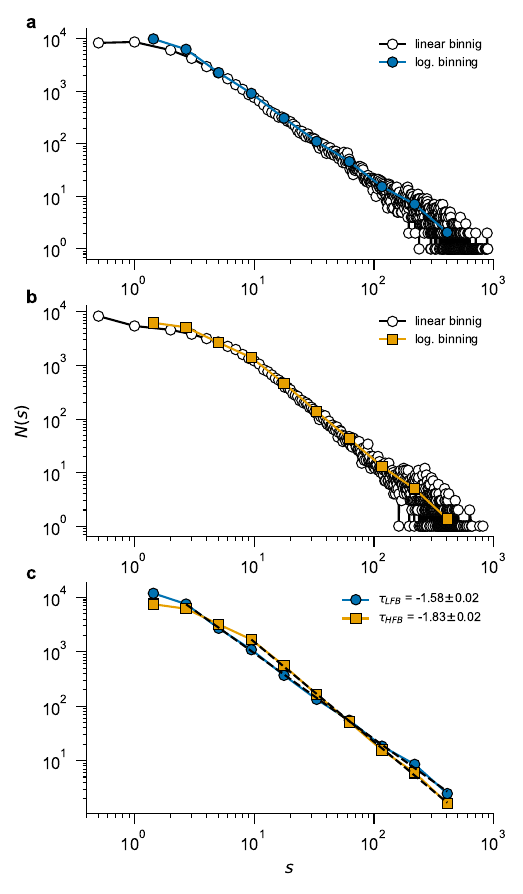}
\caption{Avalanche size distributions both in linear and logarithmic binings for (a) LFG and (b) HFG. Only the last $49,000$ events are included in the analysis. The events of $s=0$ are plotted at $s=0.5$. (c) Comparison between both distributions and calculation of exponent values, in the range of the dashed lines, resulting in $\tau = 1.58$ for LFB and $\tau = 1.83$ for HFB.}
\label{fig_dist_sizes} 
\end{figure}

\subsection{Avalanche size distributions}\label{Avadist1}

The size distributions $N(s)$ for the two experiments are shown in Fig.~\ref{fig_dist_sizes}. In both cases, the distributions span nearly three decades and follow power-law behaviors. The numbers of events with $s=0$ (i.e., beads that land on the pile without triggering an avalanche) are also comparable in the two experiments, with $N(s=0)\sim 8000$. 

As expected, increasing the friction between grains leads to an increase in the global angle of repose. Consequently, the high-friction beads (HFB) form a larger pile, with an average size of $3919$ beads compared to $3430$ beads for the low-friction beads (LFB). A larger pile means a larger reservoir of potential energy, and consequently, the possibility of generating larger events. However, the extension of the largest events are similar in both piles. In addition, the frequency of large events is smaller in the case of HFB, which indicates that the propagation of large avalanche is affected by the high dissipation linked to the increase in friction coefficient.    

\begin{figure*}[t!]
\includegraphics[width=5in]{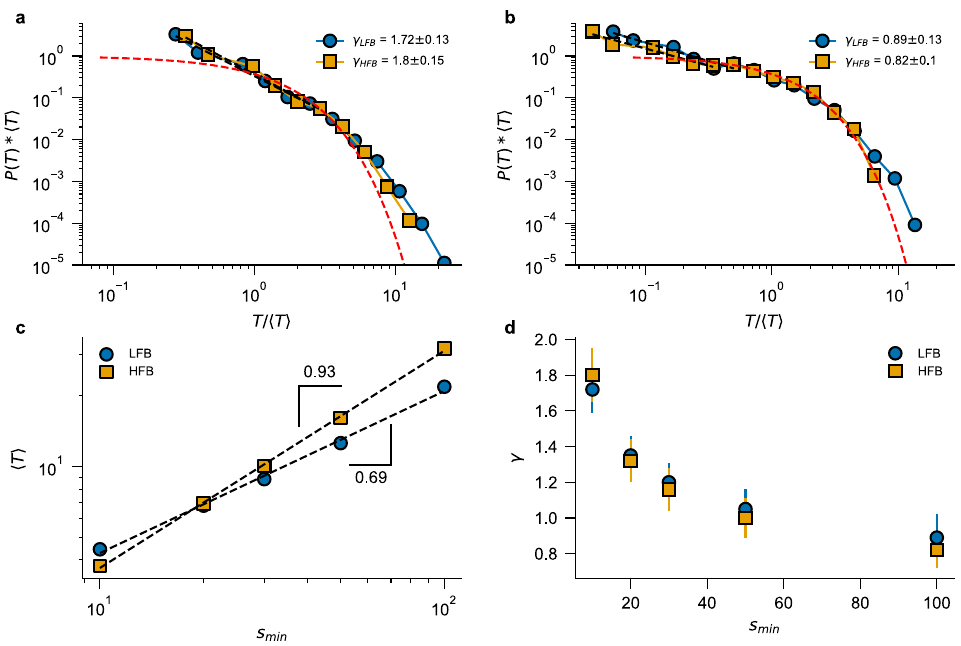}
\caption{Distributions of inter-event times $T$, normalized by the mean inter-event time $\langle T \rangle$, for both LFB and HFB, computed for avalanche sizes above a threshold: (a) $s_{\min}=10$ and (b) $s_{\min}=100$. The dashed lines indicate exponential decays. The exponents $\gamma$, extracted for $T < \langle T \rangle$, are also reported.  
(c) Mean inter-event time $\langle T \rangle$ as a function of the threshold $s_{\min}$, exhibiting a power-law dependence.  
(d) Exponent $\gamma$ as a function of $s_{\min}$, showing a decrease with increasing threshold and no significant differences between LFB and HFB.}
\label{fig_time_corr}
\end{figure*}

The avalanche exponents are $\tau = 1.58$ for LFB and $\tau = 1.83$ for HFB (Fig.~\ref{fig_dist_sizes}c). This increase in $\tau$ with friction follows the same trend observed in the OFC model for a large-dissipation regime \cite{OFC1992}: {\it increasing dissipation reduces the probability of large avalanches, which results in a larger exponent value.} The fact that the upper cutoffs ($s_{max}$) seem not very sensitive to dissipation is another similarity shared by these granular piles and the OFC model \cite{Duplat2026}.

Following the power-laws distributions down to small values we would expect a larger number of small events in the case of a larger exponent value. However, the more disordered structure observed near the apex of the HFB pile (Fig.~\ref{fig_setup}) also appears to be more stable, exhibiting less activity for small avalanches ($s<4$) than the more crystalline structure of the LFB pile. Consistently, the HFB distribution flattens for $s<8$, whereas in the LFB case the power-law regime extends down to $s=3$.

\subsection{Inter-event time distributions}\label{Timedist1} 

A key feature of this experiment is that the unit of time is set by the discrete dropping of beads onto the pile. The impacts are sufficiently gentle that no avalanche is triggered in approximately $8000$ dropping events, allowing the system to be classified as slowly driven. However, all observed activity is directly linked to structural perturbations induced by these dropping events, and no activity is detected when the driving is halted.

As a result, the driving mechanism corresponds to a stepwise increase of energy, with all activity occurring immediately after each increment. This contrasts with the smoother driving processes observed in both natural~\cite{Main1996} and laboratory faults~\cite{Lherminier2019}, where the driving does not directly perturb the dynamics beyond a gradual energy increase.

In the OFC model, inter-event time distributions exhibit Corral-like relations~\cite{corral2004}, characterized by an exponential decay for $T > \langle T \rangle$, and a power-law behavior for $T < \langle T \rangle$, where $T$ and $\langle T \rangle$ are respectively the time between two consecutive events and the mean time between consecutive events, both considering events with sizes above a given threshold $s_{\min}$. 

The corresponding exponent $\gamma$ of the power-law regime increases with dissipation \cite{Duplat2026}, indicating enhanced memory effects. This exponent is also robust with respect to variations in $s_{\min}$, a behavior observed in both seismicity~\cite{corral2004} and granular fault experiments~\cite{Lherminier2019}.

The experimental distributions of inter-event times $T$, computed for avalanche sizes above a threshold $s_{\min}$ and normalized by the mean inter-event time $\langle T \rangle$, are shown in Fig.~\ref{fig_time_corr}. They exhibit clear deviations from the exponential decay, not only for $T < \langle T \rangle$ (as in earthquake-like dynamics~\cite{OFC1992,corral2004,Lherminier2019}), but also for $T > \langle T \rangle$. In fact, two power-law regimes provide a better description of the distributions than the classical scaling proposed by Corral~\cite{corral2004}, possibly reflecting the discrete driving induced by bead deposition.

The mean inter-event time $\langle T \rangle$ [Fig.~\ref{fig_time_corr}(c)], which is directly related to the avalanche size distribution, follows the expected power-law scaling. Larger events correspond to longer waiting times, with a higher exponent observed for the HFB configuration, indicating that large events become increasingly rare. In contrast, the exponent $\gamma$ does not show significant dependence on dissipation and is not robust, but decreases with increasing $s_{\min}$. Nevertheless, for large $s_{\min}$, its value approaches that of the OFC model ($\gamma \sim 0.8$), with the tail of the distribution tending toward an exponential decay \cite{Duplat2026}.     

\subsection{Internal structure}\label{Struct1} 

\begin{figure}[t!]
\includegraphics[ width=3.4in]{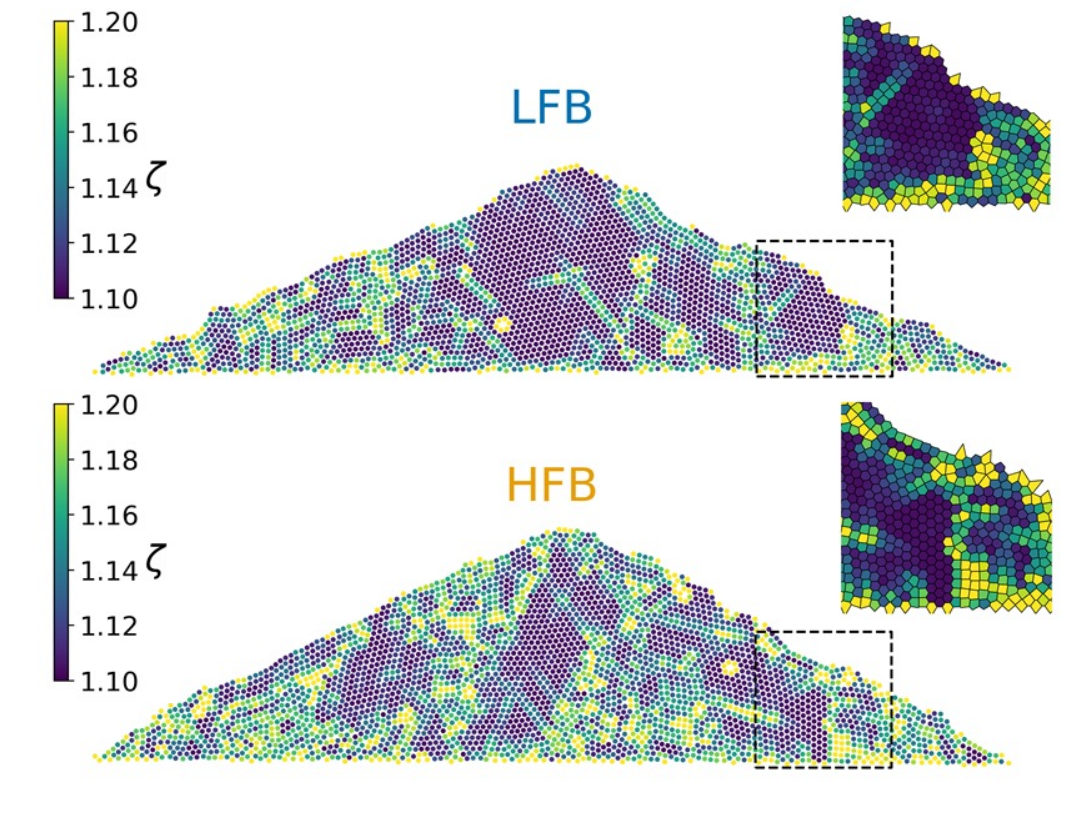}
\caption{Structure of typical piles for LFB and HFB, characterized by the shape factor $\zeta=C^2/4\pi S$, accounting for the local order of the packing. The LFB pile shows a crystalline structure at its central part, while HFB form a larger and more disordered pile.}
\label{fig_voronoi} 
\end{figure}

From the center of the beads, we have built Voronoi cells excluding the beads at the base and at the surface of the pile. The shape factor, $\zeta=C^2/4\pi S$, where $C$ is the perimeter of the cell and $S$ its area, accounts for the local disorder of the pile (Fig.~\ref{fig_voronoi}). In the case of LFB, a crystalline structure forms and evolves in the central part of the pile, with respect to a globally more disordered structure in the HFB (Fig.~\ref{fig_voronoi}). 

High friction allows the system to get locally trapped in disordered configurations. Compared to LFB, high frictional interactions dissipate more energy which stabilizes the system resulting in larger piles. Distant areas become more independent, penalizing large events. We should expect that energy is further released by small events. However, very small events are also affected by the high dissipation values, which reduces their frequency. As a result, there is an abundance of events in the range $10<s<30$ for the HFB with respect to the LFB (Fig.~\ref{fig_dist_sizes}c).    

Considering the OFC model, the increase of dissipation enhances the traces left by the avalanches, which temporally confine the dynamics into patches, allowing the evolution of the structure into favorable conditions to generate a system spanning event~\cite{Duplat2026}. In the granular piles discussed here, there are not clear patches-like structures. However, we can look for the mechanism responsible for the generation of large avalanches in the high dissipation regime of HFB. The change from a crystalline structure to a disordered one enhances the local evolution of the dynamics, playing the role of the confinement in the OFC model. Independent local dynamics at play will slowly creates structures that will evolve eventually arriving to a configuration of very high energy and global instability leading to a catastrophic event.

\begin{figure}[b!]
\includegraphics[ width=3.4in]{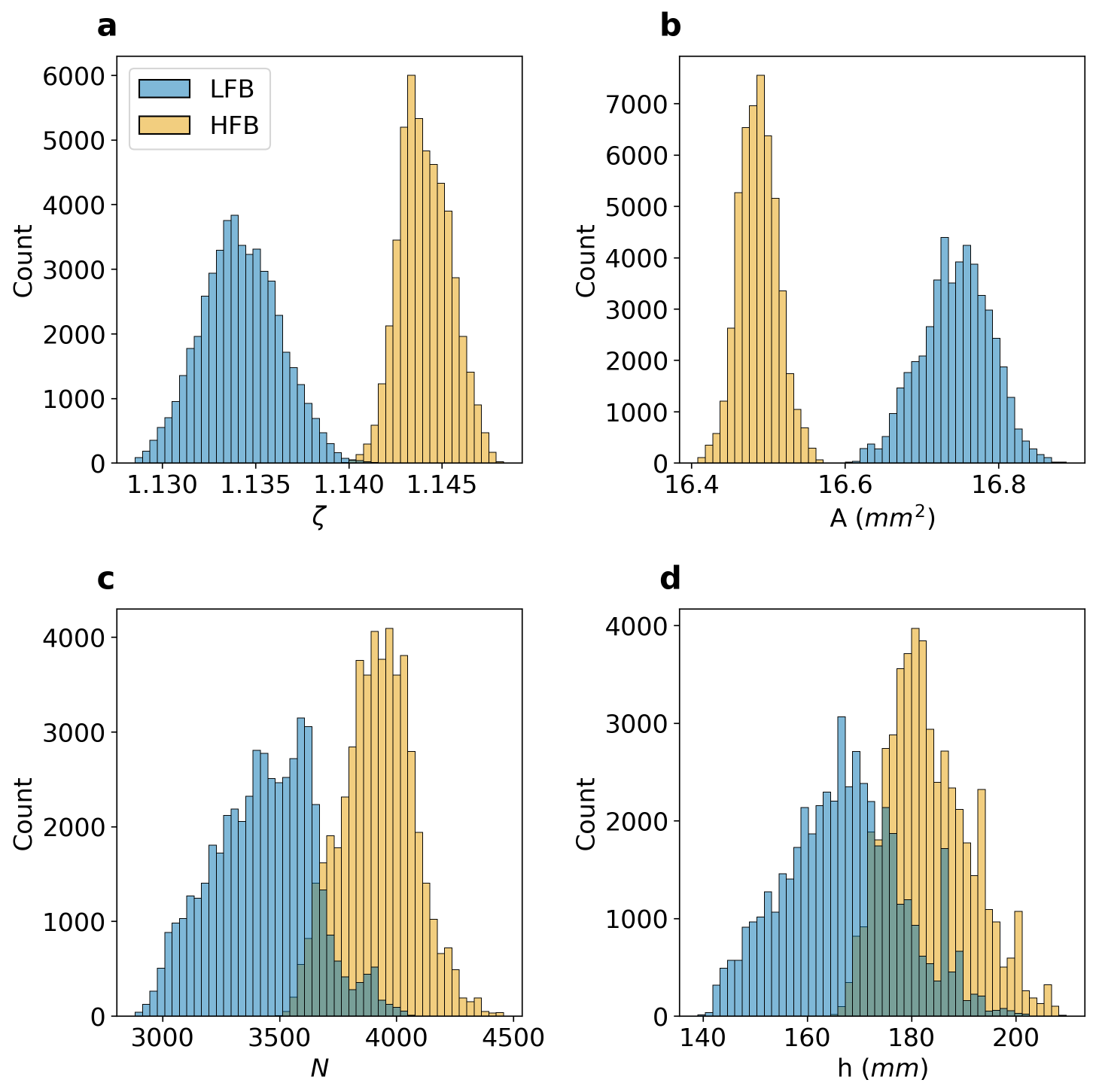}
\caption{Histograms of the main features used in the machine learning analysis: a) chape factor $\zeta$, Mean area of the Voronoi cells $A= \langle S \rangle$, c) Number of beads in the pile $N$, d) Heigh of the pile $h$.}
\label{fig_histo} 
\end{figure}

\subsection{Criticality and predictability}\label{predic1} 

Both LFB and HFB piles have similar cutoff values of about $s_{max}\sim \langle N \rangle/4$, considering $\langle N \rangle$ the average number of grains in the pile, and the increase of dissipation does not reduce the size of the cutoff, as expected in the classical mean field models of avalanches~\cite{Alstrom1988}. 

It is difficult to establish a direct analogy between the exponent values of the model and the experiment because of the different definitions of their event sizes. However, it is clear that the exponent values in the experiments are not robust (but depend on dissipation) and, at least in the HFB case, it is not universal (i.e., does not come from the application of the formalism of critical transitions that depends solely on the symmetries and dimensionality of the system \cite{LeDoussal2009}).   

\begin{figure*}[t!]
\includegraphics[ width=5.4in]{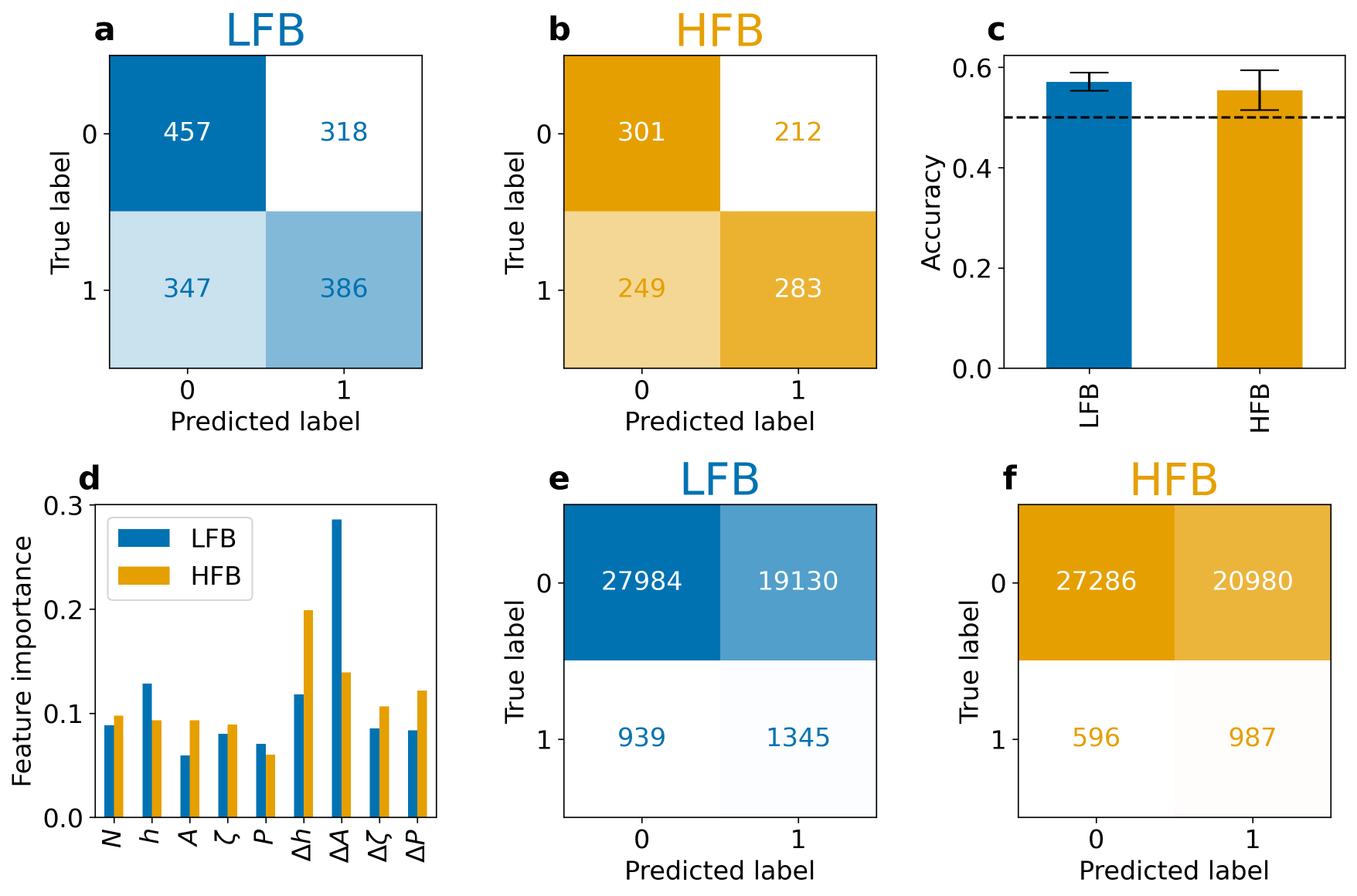}
\caption{Predicting large events. a,b) Confusion matrices in the test set containing an equal number of small and large events (balance distribution) for LFB and HFB respectively. c) Accuracy of the results, showing a ``better than random'' performance. d) Importance of the features used in the model e,f) Confusion matrices considering the whole set of data, containing the actual frequency of small and large events.}
\label{fig_pred1} 
\end{figure*}

Following recent results in the OFC model~\cite{Duplat2026}, non-robust and non-critical exponents may be related to quasi-critical dynamics where the system is not permanently critical, but hovers around the critical point, at a distance which is also power-law distributed. In a (permanent) critical system this distance is zero and the system is always ready to deliver a system-size event. As the exponents depart from its critical value, the system spends most of the time at distances far from the critical point, where it is not possible to generate a system-spanning avalanche. 

In order to test if the system is permanently or only temporarily critical, we can try to predict large events. We will consider that an event is large if $s>100$ and small otherwise. Doing so, we define a machine learning problem trying to predict the class of the next event (0 for small and 1 for large) from the structure of the pile. To represent this structure, we define a set of global features: $\zeta$ averaged over all grains, $A$ and $P$ the average surface and perimeter of the Voronoi cells, respectively, $N$ the total number of grains in the pile and $h$ the physical height of the pile. Figure~\ref{fig_histo} shows the distribution of four of these features comparing the HFP and LFB cases. As expected, we find that in HFB, the shape factor, number of grains and height of the pile are higher than in LFB, a signature of the more disordered organizations allowed by high friction. As shown in \cite{Ramos2009}, large events might be better captured by the time evolution of these features rather than their value at a given time step. Therefore, we compute four additional features representing the change in $h$, $A$, $\zeta$ and $P$ over the last 100 time steps which we write $\Delta h$, $\Delta A$, $\Delta \zeta$ and $\Delta P$.
 
The scale invariant distribution of event sizes makes large events much rarer than small ones. To avoid convergence of our models to a trivial solution where only the 0 class is predicted while maintaining high performance, we build a balanced dataset. Briefly, we ignore the first 5000 time steps and select half of all events with $s>100$. Then, we randomly select the same number of events among all of those where $s<100$ yielding 3166 events in HFP and 4568 in LFP. These are then split in a 2:1 ratio between training and testing sets. Given the relatively small amount of data points and features, we use XGBoost \cite{Chen2016}  as a good benchmark for decision tree-based models. Hyperparameters are optimized through a 10-fold validation gridsearch on the training set before being retrained on this entire set. The resulting confusion matrices on the test sets for LFB and HFB are shown in (Fig.~\ref{fig_pred1}a-b) and reach an accuracy of 59.8$\%$ and 55.9$\%$, respectively. This shows that the models are capable or distinguishing large and small events better than a random prediction, which is also visible from the average accuracy obtained over the validation set during the grid search optimization (Fig.~\ref{fig_pred1}c).

The advantage of using a tree-based model such as XGBoost is that it is interpretable, in particular we can measure the importance of each of the 9 features in the prediction (Fig.~\ref{fig_pred1}d). We find that although they are all used by the model, it focuses more heavily on time evolutions, in particular of the average surface of the Voronoi cells and the height of the pile. This effect is even more pronounced for LFB. Of note, we can obtain similar performances if we limit the features given to the model to $\Delta h$, $\Delta A$, $\Delta \zeta$ and $\Delta P$ only. 

Although the model can learn relevant differences in the structure of the pile and its evolution between small and large events, it does so on a balanced data set which does not represent the full complexity of the system. We thus test how it performs when tested against a realistic dataset. To do so, we take the pretrained models, test them on all the events that were not included in the original dataset and compute the resulting confusion matrices (Fig.~\ref{fig_pred1}e-f). They still achieve accuracy significantly above 50$\%$ (59.3$\%$ for LFB and 56.7$\%$ for HFB) but at the cost of a very poor precision given the large number of small events which are predicted as large.

\section*{Conclusions}\label{conclusions}

The OFC model is often considered anomalous because, in its non-conservative regime, it displays non-robust and non-universal avalanche-size exponents that can even exceed the mean-field limit $\tau=3/2$. Recently~\cite{Duplat2026}, these anomalous features have been used to explain experimental and earthquake dynamics exhibiting exponent values larger than $\tau=3/2$, raising the question of the extent to which meaningful analogies can be established between the model and real dissipative systems.

Here, we have shown that some of the most relevant features of the OFC model are also present in an experimental two-dimensional bead pile displaying scale-invariant dynamics~\cite{Ramos2009}. By increasing the friction between particles, and therefore the dissipation in their interactions, the exponent increases from $\tau=1.58$ to $\tau=1.83$, a substantial change~\cite{Duplat2025}, while the cutoff $s_{\max}$ remains proportional to the system size. This contrasts with the standard picture of scale-invariant dynamics, characterized by robust and universal exponent values and a cutoff that retreats with increasing dissipation, ultimately destroying criticality~\cite{Alstrom1988, Zapperi_1995, Lauritsen1996}.

Memory effects, defined here as deviations from an exponential decay of the inter-event-time distributions, constitute another similarity between the experiment and the OFC model. Notice, however, that in the model memory effects increase monotonically with dissipation, whereas this is not the case in the experiment. Local dynamics, which appears to be a key mechanism allowing the global structure to evolve toward configurations favorable to the generation of system-spanning avalanches in highly dissipative regimes, is also shared by the model and the experiment~\cite{Duplat2026}.

The fact that a na{\"{\i}}ve algorithm can achieve ``better-than-random prediction'' on a balanced dataset containing equal numbers of small and large events provides a clear indication that the experiments do not behave as permanently critical systems. However, for a scale-invariant distribution with a large exponent, the strong imbalance between small and large events implies that even a modest error rate measured on the balanced dataset becomes strongly amplified when applied to the actual distribution. As a consequence, reliable prediction of large events becomes practically impossible, due to the high number of small events predicted as large, when considering the true scale-invariant statistics rather than a balanced collection of scenarios associated with small and large events. A similar behavior is observed in the OFC model, where na{\"{\i}}ve prediction algorithms can also reach high accuracy scores~\cite{Cochet-Escartin2026}. In the model, this accuracy increases with dissipation, although this latter feature is not reproduced by our experimental results.

Taken together, these similarities suggest that OFC-like behavior should not necessarily be regarded as anomalous. Instead, the OFC model may provide a relevant framework for understanding scale-invariant dynamics in real dissipative systems characterized by avalanche-size exponents $\tau>3/2$.

\subsection*{Acknowledgments}

This work was supported by the ANR grant ANR-22-CE30-0046 and by NFR, the
Norwegian Research Council through a Petromax and a SUP grant. 



\bibliography{anr_biblio}

@ARTICLE{Baro2013,
  author = {Bar{\'o}, Jordi and Corral, {\'A}lvaro and Illa, Xavier and Planes,
	Antoni and Salje, Ekhard KH and Schranz, Wilfried and Soto-Parra,
	Daniel E and Vives, Eduard},
  title = {Statistical similarity between the compression of a porous material
	and earthquakes},
  journal = {Phys. Rev. Lett.},
  year = {2013},
  volume = {110},
  pages = {088702},
  number = {8},
  publisher = {APS}
}

@article{Stojanova2014,
 	 title = {High Frequency Monitoring Reveals Aftershocks in Subcritical Crack Growth},
 	 author = {Stojanova, M. and Santucci, S. and Vanel, L. and Ramos, O.},
 	 journal = {Phys. Rev. Lett.},
  	volume = {112},
	  pages = {115502},
 	 numpages = {5},
 	 year = {2014},
}

@ARTICLE{Gutenberg1956,
  author = {Gutenberg, Beno and Richter, Charles Francis},
  title = {Magnitude and energy of earthquakes},
  journal = {Ann. Geophys.},
  year = {1956},
  volume = {9},
  pages = {1--15},
  number = {1},
  publisher = {Istituto Nazionale di Geofisica e Vulcanologia}
}

@Article{Bares2018,
	author={Bar{\'e}s, J.
	and Dubois, A.
	and Hattali, L.
	and Dalmas, D.
	and Bonamy, D.},
	title={Aftershock sequences and seismic-like organization of acoustic events produced by a single propagating crack},
	journal={Nature Communications},
	year={2018},
	volume={9},
	number={1},
	pages={1253},
}

@article{Main1996,
author = {Main, Ian},
title = {Statistical physics, seismogenesis, and seismic hazard},
journal = {Reviews of Geophysics},
volume = {34},
number = {4},
pages = {433-462},
year = {1996},
}

@article{OFC1992,
  title = {Self-organized criticality in a continuous, nonconservative cellular automaton modeling earthquakes},
  author = {Olami, Zeev and Feder, Hans Jacob S. and Christensen, Kim},
  journal = {Phys. Rev. Lett.},
  volume = {68},
  pages = {1244--1247},
  year = {1992},
}

@article{Chessa1999,
  title = {Universality in sandpiles},
  author = {Chessa, Alessandro and Stanley, H. Eugene and Vespignani, Alessandro and Zapperi, Stefano},
  journal = {Phys. Rev. E},
  volume = {59},
  issue = {1},
  pages = {R12--R15},
  numpages = {0},
  year = {1999},
  month = {Jan},
  publisher = {American Physical Society},
  doi = {10.1103/PhysRevE.59.R12},
  url = {https://link.aps.org/doi/10.1103/PhysRevE.59.R12}
}

@article{Ramos2009,
  title = {Avalanche Prediction in a Self-Organized Pile of Beads},
  author = {Ramos, O. and Altshuler, E. and M\aa{}l\o{}y, K. J.},
  journal = {Phys. Rev. Lett.},
  volume = {102},
  pages = {078701},
  numpages = {4},
  year = {2009},
}

@article{Altshuler2001,
  title = { \href{http://journals.aps.org/prl/abstract/10.1103/PhysRevLett.86.5490}{Avalanches in One-Dimensional Piles with Different Types of Bases}},
  author = {Altshuler, E. and Ramos, O. and Mart\'{i}nez, C. and Flores, L. E. and Noda, C.},
  journal = {Phys. Rev. Lett.},
  volume = {86},
  issue = {24},
  pages = {5490--5493},
  numpages = {0},
  year = {2001},
}

@article{Ramos2006,
  title = {Quasiperiodic Events in an Earthquake Model},
  author = {Ramos, O. and Altshuler, E. and M\aa{}l\o{}y, K. J.},
  journal = {Phys. Rev. Lett.},
  volume = {96},
  pages = {098501},
  numpages = {4},
  year = {2006},
 }

@article{Ramos2010,
title = {Criticality in earthquakes. Good or bad for prediction?},
journal = {Tectonophysics },
volume = {485},
number = {1-4},
pages = {321--326},
year = {2010},
author = {O. Ramos},
}

@article {Bak1989,
author = {Bak, Per and Tang, Chao},
title = {Earthquakes as a self-organized critical phenomenon},
journal = {Journal of Geophysical Research: Solid Earth},
volume = {94},
number = {B11},
pages = {15635--15637},
year = {1989},
}

@article{Held1990,
  title = {Experimental study of critical-mass fluctuations in an evolving sandpile},
  author = {Held, G. A. and Solina, D. H. and Solina, H. and Keane, D. T. and Haag, W. J. and Horn, P. M. and Grinstein, G.},
  journal = {Phys. Rev. Lett.},
  volume = {65},
  issue = {9},
  pages = {1120--1123},
  numpages = {0},
  year = {1990},
 }

@article{Frette1996,
  author = {Frette, Vidar and Christensen, Kim and Malthe-S\o{}renssen, Anders and Feders, Jens and J\o{}ssang, Torstein and Meakin, Paul},
  title = {Avalanche dynamics in a pile of rice},
  journal = {Nature},
  year = {1996},
  volume = {379},
  pages = {49--52},
}

@article{Altshuler2004,
  title = {\textit{Colloquium}  : Experiments in vortex avalanches},
  author = {Altshuler, E. and Johansen, T. H.},
  journal = {Rev. Mod. Phys.},
  volume = {76},
  issue = {2},
  pages = {471--487},
  numpages = {0},
  year = {2004},
 }

@article{BTW1987,
  title = {Self-organized criticality: An explanation of the 1/  \textit{f}  noise},
  author = {Bak, Per and Tang, Chao and Wiesenfeld, Kurt},
  journal = {Phys. Rev. Lett.},
  volume = {59},
  issue = {4},
  pages = {381--384},
  year = {1987},
   publisher = {American Physical Society}
}

@article{Alstrom1988,
  title = {Mean-field exponents for self-organized critical phenomena},
  author = {Alstr\o{}m, Preben},
  journal = {Phys. Rev. A},
  volume = {38},
  issue = {9},
  pages = {4905--4906},
  numpages = {0},
  year = {1988}
}

@article{LeDoussal2009,
  title = {Size distributions of shocks and static avalanches from the functional renormalization group},
  author = {Le Doussal, Pierre and Wiese, Kay J\"org},
  journal = {Phys. Rev. E},
  volume = {79},
  issue = {5},
  pages = {051106},
  numpages = {34},
  year = {2009},
 }

@article {Daniels2008,
author = {Daniels, Karen E. and Hayman, Nicholas W.},
title = {Force chains in seismogenic faults visualized with photoelastic granular shear experiments},
journal = {J. Geophys. Res. Solid Earth},
volume = {113},
number = {B11},
pages = {2156--2202},
year = {2008},
}

@article{corral2004,
  title = {Long-Term Clustering, Scaling, and Universality in the Temporal Occurrence of Earthquakes},
  author = {Corral, \'Alvaro},
  journal = {Phys. Rev. Lett.},
  volume = {92},
  pages = {108501},
  numpages = {4},
  year = {2004},
}

@article{Rosendahl1993,
  title = {Persistent self-organization of sandpiles},
  author = {Rosendahl, J. and Veki\ifmmode \acute{c}\else \'{c}\fi{}, M. and Kelley, J.},
  journal = {Phys. Rev. E},
  volume = {47},
  issue = {2},
  pages = {1401--1404},
  numpages = {0},
  year = {1993},
}

@article{Maloy2006,
  title = {Local Waiting Time Fluctuations along a Randomly Pinned Crack Front},
  author = {M\aa{}l\o{}y, Knut J\o{}rgen and Santucci, St\'ephane and Schmittbuhl, Jean and Toussaint, Renaud},
  journal = {Phys. Rev. Lett.},
  volume = {96},
  pages = {045501},
  numpages = {4},
  year = {2006},
}

@article{Lherminier2019,
  title = {Continuously Sheared Granular Matter Reproduces in Detail Seismicity Laws},
  author = {Lherminier, S. and Planet, R. and Levy dit Vehel, V.  and Simon, G. and Vanel, L. and M\aa{}l\o{}y, K. J. and Ramos, O.},
  journal = {Phys. Rev. Lett.},
  volume = {122},
  pages = {218501},
  numpages = {6},
  year = {2019},
}

@article{Zadeh2019a,
  title = {Crackling to periodic dynamics in granular media},
  author = {Abed Zadeh, Aghil and Bar\'es, Jonathan and Behringer, Robert P.},
  journal = {Phys. Rev. E},
  volume = {99},
  pages = {040901},
  numpages = {6},
  year = {2019},
}

@article{Kawamura2012,
  title = {Statistical physics of fracture, friction, and earthquakes},
  author = {Kawamura, Hikaru and Hatano, Takahiro and Kato, Naoyuki and Biswas, Soumyajyoti and Chakrabarti, Bikas K.},
  journal = {Rev. Mod. Phys.},
  volume = {84},
  pages = {839--884},
  numpages = {0},
  year = {2012},
}

@article{DEARCANGELIS2016,
title = "Statistical physics approach to earthquake occurrence and forecasting",
journal = "Physics Reports",
volume = "628",
pages = "1 - 91",
year = "2016",
author = "Lucilla de Arcangelis and Cataldo Godano and Jean Robert Grasso and Eugenio Lippiello",
}

@article{Zapperi_1995,
  title = {Self-Organized Branching Processes: Mean-Field Theory for Avalanches},
  author = {Zapperi, Stefano and Lauritsen, Kent B\ae{}kgaard and Stanley, H. Eugene},
  journal = {Phys. Rev. Lett.},
  volume = {75},
  issue = {22},
  pages = {4071--4074},
  numpages = {0},
  year = {1995},
  month = {Nov},
  publisher = {American Physical Society},
  doi = {10.1103/PhysRevLett.75.4071},
  url = {https://link.aps.org/doi/10.1103/PhysRevLett.75.4071}
}

@article{Lauritsen1996,
  title = {Self-organized branching processes: Avalanche models with dissipation},
  author = {B\ae{}kgaard Lauritsen, Kent and Zapperi, Stefano and Stanley, H. Eugene},
  journal = {Phys. Rev. E},
  volume = {54},
  issue = {3},
  pages = {2483--2488},
  numpages = {0},
  year = {1996},
  month = {Sep},
  publisher = {American Physical Society},
  doi = {10.1103/PhysRevE.54.2483},
  url = {https://link.aps.org/doi/10.1103/PhysRevE.54.2483}
}

@article{Dhar_1999,
title = {The Abelian sandpile and related models},
journal = {Physica A: Statistical Mechanics and its Applications},
volume = {263},
number = {1},
pages = {4-25},
year = {1999},
note = {Proceedings of the 20th IUPAP International Conference on Statistical Physics},
issn = {0378-4371},
doi = {https://doi.org/10.1016/S0378-4371(98)00493-2},
url = {https://www.sciencedirect.com/science/article/pii/S0378437198004932},
author = {Deepak Dhar}
}

@Article{Houdoux2021,
author={Houdoux, David
and Amon, Axelle
and Marsan, David
and Weiss, J{\'e}r{\^o}me
and Crassous, J{\'e}r{\^o}me},
title={Micro-slips in an experimental granular shear band replicate the spatiotemporal characteristics of natural earthquakes},
journal={Communications Earth {\&} Environment},
year={2021},
month={May},
day={14},
volume={2},
number={1},
pages={90},
issn={2662-4435},
doi={10.1038/s43247-021-00147-1},
url={https://doi.org/10.1038/s43247-021-00147-1}
}

@misc{Bouchaud2024,
      title={The Self-Organized Criticality Paradigm in Economics \& Finance}, 
      author={Jean-Philippe Bouchaud},
      year={2024},
      eprint={2407.10284},
      archivePrefix={arXiv},
      primaryClass={q-fin.GN},
      url={https://arxiv.org/abs/2407.10284}, 
}

@article {Beggs2003,
	author = {Beggs, John M. and Plenz, Dietmar},
	title = {Neuronal Avalanches in Neocortical Circuits},
	volume = {23},
	number = {35},
	pages = {11167--11177},
	year = {2003},
	doi = {10.1523/JNEUROSCI.23-35-11167.2003},
	publisher = {Society for Neuroscience},
	issn = {0270-6474},
	journal = {Journal of Neuroscience}
}

@article{Fisher19998,
title = {Collective transport in random media: from superconductors to earthquakes},
journal = {Physics Reports},
volume = {301},
number = {1},
pages = {113-150},
year = {1998},
issn = {0370-1573},
doi = {https://doi.org/10.1016/S0370-1573(98)00008-8},
url = {https://www.sciencedirect.com/science/article/pii/S0370157398000088},
author = {Daniel S. Fisher}
}

@article{Hergarten2002,
  title = {Foreshocks and Aftershocks in the Olami-Feder-Christensen Model},
  author = {Hergarten, Stefan and Neugebauer, Horst J.},
  journal = {Phys. Rev. Lett.},
  volume = {88},
  issue = {23},
  pages = {238501},
  numpages = {4},
  year = {2002},
  month = {May},
  publisher = {American Physical Society},
  doi = {10.1103/PhysRevLett.88.238501},
  url = {https://link.aps.org/doi/10.1103/PhysRevLett.88.238501}
}

@article{Helmstetter2004,
  title = {Properties of foreshocks and aftershocks of the nonconservative self-organized critical Olami-Feder-Christensen model},
  author = {Helmstetter, Agn\`es and Hergarten, Stefan and Sornette, Didier},
  journal = {Phys. Rev. E},
  volume = {70},
  issue = {4},
  pages = {046120},
  numpages = {13},
  year = {2004},
  month = {Oct},
  publisher = {American Physical Society},
  doi = {10.1103/PhysRevE.70.046120},
  url = {https://link.aps.org/doi/10.1103/PhysRevE.70.046120}
}

@article{Peixoto_Prado2006,
  title = {Network of epicenters of the Olami-Feder-Christensen model of earthquakes},
  author = {Peixoto, Tiago P. and Prado, Carmen P. C.},
  journal = {Phys. Rev. E},
  volume = {74},
  issue = {1},
  pages = {016126},
  numpages = {9},
  year = {2006},
  month = {Jul},
  publisher = {American Physical Society},
  doi = {10.1103/PhysRevE.74.016126},
  url = {https://link.aps.org/doi/10.1103/PhysRevE.74.016126}
}

@Article{Budrikis2024,
author={Budrikis, Zoe},
title={100 years of the Ising model},
journal={Nature Reviews Physics},
year={2024},
month={Sep},
day={01},
volume={6},
number={9},
pages={530-530},
issn={2522-5820},
doi={10.1038/s42254-024-00760-x},
url={https://doi.org/10.1038/s42254-024-00760-x}
}

@article{gardner1970life,
  author  = {Gardner, Martin},
  title   = {Mathematical Games: The fantastic combinations of John Conway's new solitaire game "Life"},
  journal = {Scientific American},
  volume  = {223},
  number  = {4},
  pages   = {120--123},
  year    = {1970}
}

@article{Xu2019,
  title = {Criticality in failure under compression: Acoustic emission study of coal and charcoal with different microstructures},
  author = {Xu, Yangyang and Borrego, Angeles G. and Planes, Antoni and Ding, Xiangdong and Vives, Eduard},
  journal = {Phys. Rev. E},
  volume = {99},
  issue = {3},
  pages = {033001},
  numpages = {12},
  year = {2019},
  month = {Mar},
  publisher = {American Physical Society},
  doi = {10.1103/PhysRevE.99.033001},
  url = {https://link.aps.org/doi/10.1103/PhysRevE.99.033001}
}

@misc{Duplat2026,
      title={Memory effects govern scale-free dynamics beyond universality classes}, 
      author={K. Duplat and A. Douin and O. Ramos},
      year={2026},
      eprint={2602.00374},
      archivePrefix={arXiv},
      primaryClass={cond-mat.stat-mech},
      url={https://arxiv.org/abs/2602.00374}, 
}

@misc{Duplat2025,
      title={Gutenberg-Richter-like relations in physical systems}, 
      author={K. Duplat and G. Varas and O. Ramos},
      year={2025},
      eprint={2512.17615},
      archivePrefix={arXiv},
      primaryClass={cond-mat.stat-mech},
      url={https://arxiv.org/abs/2512.17615}, 
}

@article{Navas2019,
  title = {Universality of power-law exponents by means of maximum-likelihood estimation},
  author = {Navas-Portella, V\'{\i}ctor and Gonz\'alez, \'Alvaro and Serra, Isabel and Vives, Eduard and Corral, \'Alvaro},
  journal = {Phys. Rev. E},
  volume = {100},
  issue = {6},
  pages = {062106},
  numpages = {14},
  year = {2019},
  month = {Dec},
  publisher = {American Physical Society},
  doi = {10.1103/PhysRevE.100.062106},
  url = {https://link.aps.org/doi/10.1103/PhysRevE.100.062106}
}

@inproceedings{Chen2016,
author = {Chen, Tianqi and Guestrin, Carlos},
title = {XGBoost: A Scalable Tree Boosting System},
year = {2016},
isbn = {9781450342322},
publisher = {Association for Computing Machinery},
address = {New York, NY, USA},
url = {https://doi.org/10.1145/2939672.2939785},
doi = {10.1145/2939672.2939785},
booktitle = {Proceedings of the 22nd ACM SIGKDD International Conference on Knowledge Discovery and Data Mining},
pages = {785--794},
numpages = {10},
location = {San Francisco, California, USA},
series = {KDD '16}
}

@misc{Cochet-Escartin2026,
      title={AI prediction of extreme scale-invariant events in the OFC model}, 
      author={O. Cochet-Escartin and K. Duplat and A. Douin and  O. Ramos},
      year={2026},
      note={to be submitted},
}
\end{document}